\documentstyle[12pt,epsf,aaspp4]{article}

\newcommand{\etal}{{\it et~al.}}
\newcommand{\ie}{{\it i.e.\ }}
\newcommand{\eg}{{\it e.g.\ }}
\newcommand{\cf}{{\it cf.\ }}

\begin{document}

%
%
\title{Ultraviolet images of the gravitationally lensed quadruple quasar
Q2237+0305 with the {\it HST}
\footnote{Based on observations with the NASA/ESA {\it Hubble Space 
Telescope}, obtained at the Space Telescope Science Institute, which is 
operated by AURA under NASA contract NAS 5-26555.} 
WFPC2}
\author{Michael Blanton, Edwin L. Turner}
\affil{Princeton University Observatory, Princeton, NJ 08544 }
\affil{ blanton, elt@astro.princeton.edu}
\author{and \break Joachim Wambsganss}
\affil{Astrophysikalisches Institut Potsdam, 14482 Potsdam, Germany} 
\affil{ jwambsganss@aip.de }

%
%
\begin{abstract} 
We present and analyze observations of the quadruple lensed quasar
Q2237+0305, obtained with the {\it HST} WFPC2 camera in the F336W and F300W
filters.  Twenty-five exposures were performed within 15 hours real
time on 3 November 1995. On a timescale of 3--4 hours, we observe no
variation in component A of greater than 0.02 mag. The other
components remain constant over a period of 10 hours to within about
0.05 mag. In the final 5 hours there is some evidence (not
conclusive) for variation of component D by about 0.1 mag.  The
exposures indicate that component A is brighter than component B by
about 0.3 mag. Components C and D are fainter than component A by
about 1.3 and 1.4 mag, respectively.  Our results place an upper limit
on any fifth (central) component of 6.5 mag fainter than component A.

We determine the astrometric properties of the lens system, using only
the exposures of the higher resolution Planetary Camera chip.  We
measure the relative distances of the four components with high
accuracy. Our values are systematically larger than those of other
investigators (by 0.1\% to 2.0\%). We discuss the reasons why we
believe our results are reliable.

The F336W filter had been chosen for the observations because it
corresponds to the redshifted Ly-$\alpha$ line of the quasar.  This
filter might have allowed us to see extended Ly-$\alpha$ emission from
the Broad-Line Region (BLR) of the quasar as Ly-$\alpha$ arcs, and
hence to determine the physical size of the BLR.  However, the quasar
components in this filter are consistent with a point source.  We
conclude that there cannot be any Ly-$\alpha$ feature in the image
plane brighter than about 23.5 mag in F336W and further from the
quasar core than 100 mas. According to a lensing model by Rix,
Scheider \& Bahcall (1992), this would preclude any such features in
the source plane further than 20 mas ($\sim 100 h^{-1}$ pc, assuming
$q_0 = 0.5$) from the quasar core and brighter than 25 mag before
magnification.
\end{abstract}

%
%
\bigskip
\bigskip
 
%
%
\section{Introduction} 

The quasar Q2237+0305 at redshift $z = 1.695$ is gravitationally
lensed by a nearby galaxy at $z = 0.039$ (Huchra \etal~1985).  The
galaxy core lies nearly perfectly along the line of sight. Such
a configuration results in a symmetric, cross-like arrangement of the
four quasar component images, with relative separations in this case
between 1.2 and 1.8 arcsec.  

Several facts make this lens system useful. First, the closeness of
the lensing spiral galaxy allows us to study it in great
detail. Second, the large leverage between lens plane and source plane
that results from this proximity reduces the time scale for
microlensing. Third, the symmetric arrangement of the four components
leads to a small relative time delay (of order a day or shorter, \cf
Wambsganss \& Paczy\'nski 1994).  This last fact helps to distinguish
intrinsic fluctuations of the quasar from microlensing-induced
changes: the former have to show up in all four components almost
simultaneously, whereas the latter are completely independent of each
other, and occur on a timescale of months. Finally, the fact that the
four quasar components are bright, well separated, and of comparable
optical brightness makes Q2237+0305 an easy target for various
photometric, spectroscopic and astrometric studies.

Recently, Q2237+0305 was detected at radio wavelengths with the VLA at 3.6cm
and 20cm (Falco \etal~1996). The relative positions of the components measured
in radio agree well with the optical positions.  Furthermore, the relative flux
ratios of the components in each wavelength regime are similar, with the
exception that component D is brighter in the radio, compared with the optical
light.  Thus, the relative brightnesses in radio agree much better with the
ratios predicted by various models (Rix, Schneider \& Bahcall 1992), than do
those in the optical; the slight discrepancy in the optical is possibly caused
by a demagnification due to microlensing (\cf Wambsganss, Paczy\'nski \&
Schneider 1990; Witt \& Mao 1994), or by dust in the lensing galaxy.

The system Q2237+0305 was the first in which microlensing was detected
(Irwin \etal~1989; Corrigan \etal~1991). There are various
photometric monitoring programs underway; these continue to show
fluctuations in the component intensities which can be attributed to
microlensing.  Recent results of some of these campaigns were
published by Lewis, Irwin \& Hewett~(1995), and Ostensen \etal~(1996).

Here we present new data obtained with the WFPC2 camera aboard the
Hubble Space Telescope. The motivation and the technical details of
the observations are explained in Section 2.  In Section 3 we present
the results of these observations, including the photometry and
the astrometry of the four components.  We discuss our results and their
implications in the final Section 4.

%
%
\section{Motivation and Observations}

The primary goal of the proposed observations was to use the
Q2237+0305 lens system as a ``Zwicky telescope,'' \ie to take
advantage of the lensing magnification which nature provides in this
case to image the background QSO's structure with unprecedented and
otherwise unavailable resolution.  

We chose Q2237+0305 in part because Yee \& De Robertis (1991) had
claimed to see Ly-$\alpha$ emission distinct from the quasar
components, possibly resulting from an intense star-forming region
associated with the quasar.  In addition, the quasar's distance
redshifts the Ly-$\alpha$ line to 3270\AA, which is available to HST
but is not so red that the light of the lensing galaxy swamps the
quasar image. High surface brightness, spatially extended structure in
the quasar image will most likely appear in Ly-$\alpha$. Such
structure could result from the broad line region (BLR), from a larger
and more quiescent quasar HII region, or possibly even from intense
star formation regions in the host galaxy. Also, if we were so
fortunate as to resolve the continuum emitting region in some way, it
is also expected to be relatively bright in the UV. Meanwhile, the
light of the relatively red bulge component of the lensing galaxy,
through which we see the quasar components, provides little UV
background to swamp low surface brightness parts of the source image.

We proposed obtaining deep WFPC2 images in the F336W and F300W bands.
Both bands are near the Ly-$\alpha$ line, but Ly-$\alpha$ would appear
brighter in F336W than F300W. This difference allows us to
distinguish between continuum and Ly-$\alpha$ emission.

We hoped, in the most optimistic scenario, that this project could
yield the first well-resolved UV image of a classical high redshift
QSO and thus provide a fundamentally important datum in our attempts
to understand quasars.  A Ly-$\alpha$ ring image, similar to the
radio ``Einstein rings", was an extreme but real possibility.
Finally, even if no resolved images were detected, the data would
provide useful upper bounds on sizes and surface brightnesses of
structures near the core of this QSO and so are still of considerable
interest.

On 3 November 1995, we obtained a total of twenty-five exposures
using the WFPC2 camera aboard the Hubble Space Telescope. Ten of these
images were exposed on one of the Wide Field chips; of these images,
four were taken using the F300W filter and six were taken using the
F336W filter. The Wide Field images were exposed for approximately
1200~sec each. The remaining fifteen images were exposed on the
Planetary Camera chip; of these images, six were taken using the F300W
filter and nine were taken using the F336W filter. The Planetary
Camera images were exposed for approximately 800~sec each. Thus, we had
four combinations of chip and filter. In addition, within each of
these combinations, the images were dithered by approximately
half-integer numbers of pixels (\eg $2\frac{1}{2}$ or
$5\frac{1}{2}$). This dithering procedure allowed us to recover
resolution higher than that of the chips' pixels. See Table 1 for
details on all of these exposures. 

The PC chip scale is $45.53\pm 0.02$ mas/pixel, and the scale of the
WF3 chip is $99.56\pm 0.04$ mas/pixel (Holtzmann \etal~1995).  Thus,
the PC achieves finer resolution while the WF3 yields greater
signal-to-noise. We would expect the WF3 chip to detect large and
diffuse signals more readily than the PC chip, and the PC chip to
better perceive features close to the quasar core.  Additionally, the
PC chip is more suited to astrometry than the WF3 chip. Because of
contamination difficulties, the PC chip is also preferable for
photometry.

For the purpose of some of the work described here, it was convenient
to combine the various images. This combination required using the
{\it IRAF}\footnote{{\it IRAF} is distributed by the National Optical
Astronomy Observatories, which are operated by the Association of
Universities for Research in Astronomy, Inc., under cooperative
agreement with the National Science Foundation.} routine {\tt crrej}
to remove cosmic rays, and the {\it IRAF} routine {\tt drizzle} to
account for the dithering of the images (Fruchter \& Hook 1997). Thus,
we obtained four final images: two filters (F300W and F336W) times two
chips (PC and WF3). These four images appear in Figure 1.

Finally, we note that the field around Q2237+0305 is extremely empty
in our UV images. Only two field stars of significant brightness were
detected in our exposures, both of them on exposures 11--25 of Table
1, when the quasar appeared on the PC chip. One was $\sim$ 17 mag and
appeared on WF2; the other was $\sim$ 19 mag and appeared on WF3. This
dearth of field stars, and their absence on the PC chip, posed
difficulties in the photometric and astrometric calibration.

%
%
\section{Results}

As well as addressing the issues related to the physics of quasars
outlined in Section 2, these exposures provide excellent astrometric
and good photometric information. Thus, before discussing the
possibility of resolving a Ly-$\alpha$ emitting region in the lensed
quasar, we describe our photometric and astrometric results. As we
outline below, we obtained most of these results from the uncombined
images.  In order to obtain the position of the galaxy core, however,
it was necessary to use the combined images.  Finally, we examine the
possibility of a resolvable BLR or continuum-emitting region, using
the combined images to obtain maximum signal-to-noise.

%
%
\subsection{Photometry}

We measure the relative brightness of the four quasar components A, B,
C, D (identified in Figure 1) in each of the twenty-five exposures,
using the {\it IRAF} procedure {\tt qphot}.  The zeropoint magnitudes
for the STMAG system used here (the constant flux system used by Space
Telescope) were derived from the {\tt PHOTFLAM} values associated with
each filter and chip. The errors in the zeropoints are about 0.02 mag
(Holtzmann \etal~1995). The other major source of error is
contamination, which accumulates on the WFPC2 after each monthly
decontamination and degrades the throughput of the instrument. The
effect of contamination is more severe for filters of shorter
wavelength, and varies between the chips. For F300W and F336W, it is
about 0.01 -- 0.05 mag. Following Whitmore, Heyer \& Baggett~(1996),
we used the {\tt synphot} package and a synthetic quasar spectrum with
$z=1.695$ to estimate the effect of contamination in our case. Our
results yield about 0.01--0.02 mag contamination in the PC chip, and
0.04--0.06 mag contamination in the WF3 chip. Because the
contamination corrections are rather uncertain, we use only the PC
chip to obtain the overall photometry. We average our results for each
filter in the PC chip and list them in Table 2. The errors listed
there are simply the $1\sigma$ variance found in the determination of
the mean, added in quadrature with the errors expected in the
zeropoint and in the contamination correction ($\sim$ 0.01 mag in
F336W, $\sim$ 0.02 mag in F300W).

In Figures 2a and 2b, we examine the time dependence of the component
brightnesses by plotting the photometry for each image of Table
1. Most of the exposures of the A component on WF3 have saturated
pixels and are therefore not included in the plot. Other exposures are
not included because cosmic rays interfere with the determination of
their photometry. 

It is useful to examine two time scales. First, because of the contamination
and zeropoint uncertainties, we examine each chip/filter combination
separately. This examination probes a timescale of only 3--4 hours, but allows
us to assume that zeropoint and contamination problems are uniform for a single
chip/filter combination.  Accordingly, the error bars on the data points in
Figures 2a and 2b do not include the zeropoint and contamination errors, but
only the statistical photometric errors convolved with 0.01 mag pixel-centering
errors. To test for variation, we compare the measured $1\sigma$
($\sigma_{\mathrm{obs}}$) dispersion about the mean (dotted lines Figures 2a
and 2b) with the estimated $1\sigma$ ($\sigma{\mathrm{est}}$) errors on the
data points. If the ratio $\sigma_{\rm obs}/\sigma_{\rm est}$ of these two
quantities were large, we could claim possible variation. For comparison, for
our 17 mag field star on WF2 $\sigma_{\rm obs}/\sigma_{\rm est} \sim 1.5$. For
the quasar, in the PC chip, $\sigma_{\rm obs}/\sigma_{\rm est}$ exceeds 1 only
for the B component in the F336W filter; this exception is due to a single
anomalously bright image of B which we do not believe is intrinsic
variation. In the WF3 chip, the situation is more difficult, especially in the
F336W filter, where for the B, C, and D components $\sigma_{\rm
obs}/\sigma_{\rm est}\sim 2$. Excepting these final images, we state that over
the first 10 hours component A remains constant to within 0.02 mag and the
other components remain constant to within 0.05 mag.

To examine the situation in the last 5 hours more thoroughly, we plot D$-$B and
C$-$B magnitude differences in F336W versus time in Figure 3. The error bars
and dotted lines have the same meanings as in Figure 2. We notice that while C
and B remain at constant relative brightness throughout the run, and between
the PC chip to the WF3 chip, D brightens relative to B and C in the final three
images. Namely D$-$B changes by about 0.1 mag over the course of about 1.5
hrs. It is possible that this change is due to either intrinsic variability of
the quasar or microlensing.  We note that it is more likely that D brightens
than that B and C dim. First, an explanation due to microlensing would clearly
require two unlikely coincidences: that images B and C would start to vary at
the same time and by the same amount; second, since the estimates of the time
delay between B and C from lens models are typically $\sim$ 10 hours (Rix
\etal~1992), intrinsic variation of the quasar source would not cause them to
vary together either. It is worth noting that the variations of B, C, and D,
taken individually, are not excessive in these last three images ($\sigma_{\rm
obs}/\sigma_{\rm est}\sim 2$ compared to 1.5 for a field star). Therefore, we
cannot exclude that the large change in D$-$B is simply a coincidence between
the errors in components B, C, and D; that is, that B and C just happen to
appear somewhat dim in these images and that D just happens to appear to
brighten.

To investigate the second interesting time scale, we look at the difference
between the photometry on the PC chip and on the WF3 chip in each filter; this
examination extends the timescale to $\sim 8$ hours for F300W and $\sim 15$
hours for F336W, but increases our errors because we have to account for the
zeropoint and contamination difficulties. In fact, a glance at Figures 2a and
2b show that there must indeed be large contamination correction or zeropoint
errors, as there are shifts in the apparent magnitudes of up to 0.1 mag halfway
through the run, when we changed to the WF3 chip. Thus, although the PC and WF3
magnitudes are separated by $3\sigma$ even when we account for contamination
correction and zeropoint errors, we do not regard these differences as
intrinsic. (Note that any zeropoint and contamination errors do not affect our
above conclusions regarding B$-$D).

Photometry in the F336W filter was performed by Rix
\etal~(1992). Their results agreed with ours for components A and D,
but they found components B and C to be significantly brighter than we
did (about 0.4 magnitudes each). Note that the color differences
between components (presumably due to reddening by the galaxy core of
the C and D components) preclude a meaningful comparison of our
relative brightnesses to those observed in bands other than F336W or
F300W.

%
%
\subsection{Astrometry}

Since the PC chip is better suited to astrometry than the WF3 chip, we
measure the quasar and galaxy positions using only the PC chip. For
each of the fifteen exposures on this chip, we determine the position
of each quasar component ($A$, $B$, $C$, and $D$) using the {\it IRAF}
procedures {\tt imcentroid} and {\tt metric}. Because the foreground
galaxy has such a weak signal in our bands, we neglect the effect of
any gradient in the galaxy light distribution on our calculated
positions.

Because there are small rotations of the camera from exposure to
exposure, we want to define the quasar geometry in a way which does
not depend on the (slightly uncertain) orientation of the
camera. Thus, we pick five quasar component pairs ($AB$, $CB$, $DB$,
$AC$, and $DC$) and for each exposure measure the distances between
the members of each pair.  Thus, after measuring the positions of $A$,
$B$, and $C$ in an exposure, we can calculate the distances $AB$,
$BC$, and $AC$. Even if the camera orientation changes from image to
image, these distances should remain constant, as they are invariant
under rotation. Indeed, they are quite constant, and we are able to
average their values over all the exposures. These distances are
listed in Table 3. Note that the standard deviations are all
approximately 1.5 mas. When we perform the same procedure on the WF3
images, they yield results which are consistent with those of the PC
chip, but with standard deviations of order 10 mas.

Since the galaxy is quite dim in the F300W and F336W filters, it is
necessary to measure the position of its core using coadded images. We
coadd the F300W exposures and F336W exposures separately, measure the
distance of the galaxy core from components $B$ and $D$ in both
coadded images, and then average the results to obtain the distances
$BG$ and $DG$ listed in Table 3. Such a measurement is not as accurate
as the distance determinations between the (point-like) quasar
components, since the images undergo at least one shifting operation
in the process of being coadded, in addition to being subjected to
cosmic-ray rejection algorithms.  Furthermore, the galaxy core is not
point-like, but rather extended. Thus there could be systematic errors
in the $BG$ and $DG$ distances which are larger than the statistical
errors listed.

We use these distances to compare our results to those of other
investigators. Essentially, the distances provide a more robust
comparison of results than coordinates do because of the rotation
problem described above; using distances instead of coordinates
eliminates the need to worry about whether somebody else's coordinate
system is exactly lined up with ours. In rows two through six of Table
3, we have listed the results of a number of other investigators.

Immediately it is clear that our results differ systematically from
most of the previous results. In particular, our distances are between
0.1 -- 2.0 \% larger than those of all other investigators. Only for
the $BG$ distance is our result smaller. This observation raises the
possibility that we have determined our absolute scale inaccurately.
Nevertheless, because the WFPC2 is not limited by atmospheric seeing,
because it is a high quality post-repair {\it HST} observation, and
because it is in a band in which the galaxy is faint, we are still
confident of our results. The deflection angles and hence image
separations in a simple gravitational lens system are proportional to
the lens mass enclosed by the images; thus, our result implies an
increase of 0.2 -- 4.0\% for this quantity, assuming a singular
isothermal sphere model.  This factor would be negligibly small
compared to uncertainties in almost any other system, but detailed
modeling of Q2237+0305 yields a mass for the inner region of the
lensing galaxy which is thought to be accurate at such a level (Rix
\etal~1992).

Here we discuss possible reasons for the systematic discrepancies
between our astrometric results and those of others. Yee (1988), whose
values are consistently about $1.5\sigma$ below ours, measured his
astrometry in wavelengths ranging from 495 nm upwards.  It is possible
that the shape of the quasar region that dominates at these higher
wavelengths is different than the shape of the region that dominates
around 300 nm; such a shape difference could cause a slightly
different image configuration. Probably a more important effect is
that in the higher wavelengths, the foreground galaxy is bright, and
the gradient of its light distribution can skew the determined
positions of the quasar components. Of course, he accounted for this
effect using an iterative method which fit for the galaxy light
distribution and the quasar component positions together. However,
this procedure demands an accurate knowledge of the PSFs, and can be
unreliable. The same comments apply to the results of Irwin (1989) and
Racine (1991); furthermore, Racine (1991) determined his plate scale
by scaling his results to those of Yee (1988), and therefore did not
produce an independent result.  Finally, Rix \etal~(1992) and Crane
\etal~(1991) both suffer from the fact that the astrometry was
performed with a pre-repair {\it HST} instrument.  It is worth noting,
however, that the Rix \etal~(1992) results do agree with ours within
their $1\sigma$ error bars, though their individual values are
consistently smaller than ours.

Ideally, one would prefer to approach this question of scale by
measuring the angular distance to a field star from the quasar
components. Unfortunately, there are no stars on the PC chip in our
exposures bright enough to yield reliable astrometry. There do exist
two field stars that appear in the WF2 and WF3 chips during our PC
exposures. However, even though the {\it IRAF} {\tt metric} routine
will yield astrometry for these relative to the quasar in the PC chip,
such measurements would not yield unambiguous information about the
scale of the PC chip, since the WF3 scale would also be included in
the measurement. In addition, most of the other investigators either
had to mask the relevant comparison stars because of their brightness
(Racine 1991), did not have it in their field of view (Crane
\etal,~1991), or had our problem, that the stellar image was on a
different chip than the quasar (Rix \etal~1992). So measuring such an
angular distance would not help to compare our results with other
measurements.

Although for the purpose of comparing datasets it is preferable to
measure distances, for modeling the system it is more useful to have
the positions given in a particular coordinate system. Therefore, we
also have listed in Table 2 the right ascension and declination
displacements relative to component A, in units of arcsec. Note
that the errors are the same as the errors in the distances of Table
3; that is, we do not include the errors incurred because of the
uncertainty in determining the coordinate system.

%
%
\subsection{Limits on the Size of the BLR of the quasar }

The main purpose of these observations was to place limits on or,
with luck, detect the size and brightness of the broad-line region
(BLR) possibly associated with the quasar. The idea was that even
if the unlensed BLR of the quasar were slightly smaller than the pixel
size and its brightness below the detection limit, the BLR would be
stretched and magnified due to the lensing magnification, so that one
could possibly see arc-like structures. For this purpose it was
best to use the combined images; however, one must temper any
enthusiasm for the resulting signal-to-noise and resolution gains by
the awareness of possible spurious signals created by the combination
process.

We employed two methods to examine possible tangential extension of
the broad-line region. The first consisted of a mostly qualitative
examination of quasar components and their colors; the second
consisted of a quantitative examination of what sort of signal we
could possibly detect around the quasars.

Upon examining the images visually, it seems clear that there is no
measurable tangential extension. This absence is evident in Figure 4,
which depicts a sum of all the components for the F336W exposure in
the PC chip. In an attempt to detect a tangentially extended component
in the image, we rotated our image of each component into a coordinate
system such that the center of the galaxy was along the horizontal to
the left, and the clockwise direction tangent to the galaxy was
vertically down. Then we combined all of the images, scaling each to
the same magnitude before combination. Presumably, such a depiction
would enhance any tangential extension, which would appear vertically
in Figure 4. The image here is nearly circularly symmetric, but the
contours are $\sim 5\%$ longer in the vertical direction than in the
horizontal direction. Similar images created for the other filter/chip
combinations showed approximately the same result. We know that there
is asymmetry in the PSF; however, since we rotate the image of each
component before summing, we expect that this asymmetry is suppressed
to some degree. To evaluate the completeness of this suppression, we
perform the same rotation and summation, using as the base image a
field star of about 17 mag on the WF2 chip (thus, instead of starting
from four separate images we start each rotation from the same
image). In the F336W filter, the resulting asymmetry amounted to
$\sim$~5\% excess in the vertical direction; in the F300W filter, the
asymmetry was 2--3\%. We also performed the rotation and summation on
synthetic PSFs produced by the program {\tt tinytim} (Krist~1992). The
contours determined for these combined images are also long in the
vertical direction, by $\sim$~2--4\%. Therefore, we conclude that the
slight ellipticity observed in Figure 4 cannot be considered a
reliable signature of structure.

In addition to searching in general for tangential extension, we can
search specifically for features due to Ly-$\alpha$ emission. Since
F336W would transmit the red-shifted Ly-$\alpha$ emission
approximately three to four times more efficiently than would the
F300W filter, we can difference the two filters in an attempt to pick
out such features. Because the images are brighter in general in the
F336W filter, we found it best to scale the images accordingly before
differencing; this scaling was designed to make the cores of the
components cancel, so one could pick out features surrounding the
core. As the components differ somewhat in color (C and D being redder
than A and B), we performed this scaling and subtraction separately
for each component.

Figure 5 depicts this subtraction for the images in both chips. Note
that in the PC exposures, the A and B components reveal a slight
positive feature downwards and leftwards of each, at an angle of
forty-five degrees from the negative-x axis and at a distance from the
core of approximately 0.1 arcsec. There are at least three
explanations of this feature to explore. First, it could be an image
of a real BLR associated with the quasar. Second, it could be a result
of the dithering process. Finally, it could result from a difference
in the diffraction patterns for the two filters. We favor this third
explanation.

If the features were images of a Ly-$\alpha$ region associated with
the quasar, the region would have about $10^{-3}$ of the integrated
brightness of the quasar. This explanation of the features could be
confirmed if we could determine whether the features existed around
the C and D components, as well; since these components are expected
to have the opposite parity of A and B, a feature due to lensing would
appear on the opposite side in C and D than in A and B. That is, a
lensed feature appearing on the lower left in the A component would
appear on the upper right of the C component. Unfortunately, the noise
around the C and D components is too severe to allow such a test.

However, there is another property we would expect if the feature were
due to a Ly-$\alpha$ region near the quasar. We would guess that the
features would be oriented differently with respect to their
associated components. In particular, we would guess that each feature
would lie in a direction nearly perpendicular to the line connecting
its associated component and the galaxy.  The direction towards the
galaxy at A differs from that at B by thirty degrees, so we would
expect the line connecting the component and the feature at A to
differ from that at B by a similar amount. However, both of the
observed features are forty-five degrees from the x-axis, which argues
against their interpretation as a BLR.

Since the dithering occurs along the same direction (forty-five
degrees from the x-axis), we suspected that perhaps a problem with
accounting for the dither had caused the feature.  Thus, we
experimented with putting artificial errors in the dithering shift;
that is, we purposely misaligned the images before recombining
them. If the feature had been caused by an error in the dithering, we
would expect the misalignments to affect the feature. Although we
tried misalignments as drastic as half a pixel, they had no effect on
the feature. Therefore, we conclude that the dithering process most
likely did not cause the feature.

The remaining explanation is that the diffraction patterns for the two
filters differ enough that their difference causes a visible residual.
As there are no bright stars in our exposures on the PC chip with
which to test this proposition, we resort to the synthetic PSFs
provided by the program {\tt tinytim}. We subtract synthetic PSFs in
the F336W and F330W filters (subsampled by a factor of two, as the
image in Figure 5 is). Because the diffraction ring patterns are of
slightly different sizes in the two filters, one obtains a pattern of
rings in the differenced image.  Examining the first positive ring
from the center, we discovered that it was about the same distance
from the center and about the same brightness as the features in the
real images; interestingly, it had an asymmetry that caused it to be
brighter on the lower left than elsewhere along the ring. We avoid
taking this result too literally, but we do conclude that the
differing diffraction patterns of the two filters could cause the
feature we see.

Even so, we can use this feature to place an upper limit on the
brightness of compact regions that could exist without our detecting
them.  For the A component, the accompanying feature has an integrated
brightness about 6.9 magnitudes dimmer than the component; the
corresponding figure for the B component is about 7.5 magnitudes. Thus
we can state with confidence that there is no Ly-$\alpha$ feature
further than 100 mas from the quasar core and brighter than about 23.5
magnitudes in the F336W band. Using the lensing model of Rix
\etal~(1992), this would preclude any features further than 20 mas
($\sim 100 h^{-1}$ pc, assuming $q_0 = 0.5$) from the quasar core in
the source plane and brighter than 25 mag before magnification.

We can place another quantitative limit on the size and brightness of
a tangentially extended image surrounding the quasar components. To do
so, consider one of the quasar components. Draw a circle centered on
it. Evaluate the F336W surface brightness at each point around that
circle. This evaluation yields brightness as a function of angle. If
the coordinates are such that the galaxy is in the direction $\theta =
0$, and the quasar component were tangentially extended, the
brightness function would have bumps at $\theta = +90$ and $-90$
degrees. Upon examination of such plots, we see no such bumps.  To
place a limit on the size of bump we could measure, we calculate the
standard deviation $\sigma$ of the surface brightness values around
the circle about their mean. Then we require some signal-to-noise
ratio (S/N) for a meaningful result, and our limit becomes
(S/N)$\sigma$.

Figure 6 reveals the results of this procedure for the images in the
PC chip and the F336W filter, which is most likely to reveal
Ly-$\alpha$ structure. The horizontal axis is the radius in mas, and
the vertical axis is the standard deviation in flux units. Depending
on the assumed lensing model, we may divide the radius by the
appropriate magnification for each component to obtain the limits
appropriate for the source plane; thus, we can use the lensing
magnification to ``squeeze'' the plot radially. For example, a limit
of $3\times 10^{-17}$ erg cm$^{-2}$ s$^{-1}$ arcsec$^{-2}$ at 100 mas,
given a magnification of 5, would become a limit of $3\times 10^{-17}$
erg cm$^{-2}$ s$^{-1}$ arcsec$^{-2}$ at 20 mas.  However, we are
obliged to multiply the vertical axis by our desired signal-to-noise
ratio; thus we must slide the plot up (on a logarithmic axis).  In the
example above, our limit of $3\times 10^{-17}$ erg cm$^{-2}$ s$^{-1}$
arcsec$^{-2}$ at 20 mas, given a required (S/N) of 5, would yield, a
limit of $1.5\times 10^{-16}$ erg cm$^{-2}$ s$^{-1}$ arcsec$^{-2}$ at
20 mas. The results of such a squeezing and sliding appear in
Figure 7 for the model 2a of Rix \etal~(1992), with a signal-to-noise
ratio of 5.

We compare our two methods by noting that the brightest pixels in the
features associated with the A and B components have a surface
brightness of about $3\times 10^{-17}$ erg cm$^{-2}$ s$^{-1}$
arcsec$^{-2}$ and are located about 100 mas from each component. This
position is indicated as a cross in Figure 6. Thus, the feature is
just at our calculated threshold.  With the conservative requirement
of a 5$\sigma$ result, as in Figure 7, this feature would be well
below the noise. In Figure 7, we have included the source plane
position of the feature on every plot; for this purpose we used its
position as determined by its appearance next to the B component. That
this feature is somewhat submerged in our predicted noise encourages
our belief that Figure 6 provides a reasonable upper limit for the
brightness of other possible features.

Finally, we can compare our results with other observations of regions
surrounding high-redshift quasars, in particular those of Bremer
\etal~(1992). These investigators found extended Ly-$\alpha$
emission around two $z=3.6$ quasars, at fluxes of approximately
$10^{-16}$ erg cm$^{-2}$ s$^{-1}$ and extensions of about 4
arcsec. From our results, any emission at this level over such a large
scale can be excluded for quasar Q2237+0305.

%
%
\subsection{Limits on a Fifth Image}

A final consideration is the possibility of detecting a faint fifth
component of the quasar. Predictions of the brightness of this
component relative to A are uncertain by orders of magnitude; for
example, the results of Schneider \etal~(1988) indicate $\Delta m >
3.5$, Kent \& Falco (1988) find $\Delta m \approx 7.5$, and Rix,
Schneider \& Bahcall (1992) find $Delta m > 5$ for all their
models. In this regard, it is worth noting that the galaxy core is 6.5
mag dimmer than component A, and is close to the threshold of
detectability. Since there is not the slightest indication of a fifth
component in the image, it must be at least 6.5 mag dimmer than
component A, in our filters F336W and F300W.  Our results do not
appear to support the findings of Racine (1991), who detected a faint
fifth component 4.5 mag dimmer than A in the R and I bands.  However,
a central fifth component could be brighter with respect to A in
longer wavelength bands as a result of reddening due to dust. The
fifth component would shine through the very center of the lensing
galaxy, making any extinction estimates wildly uncertain.

%
%
\section{Summary and Conclusions}

In this paper, we have presented the results of Hubble Space Telescope
exposures on the WFPC2 camera of quasar Q2237+0305; specifications of
the images are in Table 1 and we display the image configuration in
Figure 1. The F336W and F300W filters were chosen because the
relatively red galaxy bulge through which we see the quasar components
is faint in the UV, and because one of the filters is very close to
the redshifted Ly-$\alpha$ line of the quasar (at about 3270\AA),
so that it is plausible that any spatially extended structures in the
quasar environment would emit in this wavelength regime. We have
determined:
\begin{enumerate}
\item The photometry for the four components in the F336W and F300W
bands, which is listed in Table 2. The relative brightnesses of the
components are known to vary with time due to microlensing. At the
time of our observations, component A was the brightest, followed by
component B which was fainter by 0.3 mag. Components C and D were
fainter than component A by about 1.3 and 1.4 mag, respectively.  On
the timescale of 3--4 hours, we can state that we observe no variation
in component A of greater than 0.02 mag. For the other components,
over a period of 10 hours they remain constant to within about 0.05
mag. In the final five hours there is some evidence for variation of
component D of about 0.1 mag.
\item The astrometry of the four components and the galaxy core, which
is listed in Tables 3 and 4. The (1$\sigma$-) uncertainty of these
astrometric measurements are about 1.5 mas. We found the
system to be consistently larger (by between 0.1\% and 2\%) than previous
studies have found. We discuss above why we think our values are
reliable.
\item The existence of a feature near the A and B components which is
bright in F336W, as shown in Figure 5 with a subtraction of F300W from
F336W. We conclude that this signal is probably an artifact of the
differing diffraction patterns of the two filters. However, we use the
brightness of the feature to obtain an upper limit on the brightness
of any real Ly-$\alpha$ regions, plotted as the crosses on Figures 6
and 7.
\item An estimate of the upper limits these images yield for the
brightness of any extended image near the quasar, as a function of
angular distance from the quasar in the source plane. These upper
limits depend on the lensing model and the desired
signal-to-noise. They are plotted in Figures 6 and 7. 
\item An upper limit on the brightness of central fifth component in
our band which is 6.5 mag fainter than component A. 
\end{enumerate}

\section{Acknowledgements}

We wish to thank Ray Lucas, Jean Surdej, and Brad Whitmore of the
Space Telescope Science Institute for useful help and advice. In
addition, we would like to thank the authors of the {\tt drizzle}
algorithm, Andy Fruchter of STScI and Richard Hook of ST-ECF. We
acknowledge financial support from STScI grant GO-05937.01-94A.A01. 

%
%

%
%

\clearpage
\begin{deluxetable}{ccccc}
\footnotesize
\tablecaption{Log of observations taken 3 Nov. 1995.} 
\tablewidth{0pt}
\tablehead{\colhead{Exposure Number} & \colhead{Camera} & 
\colhead{Filter} &
\colhead{Exposure Time (s)} & \colhead{Dither Offset (mas)}}
\startdata
1 & WF3 & F300W & 1100 & 0\nl
2 & WF3 & F300W & 1100 & 0\nl
3 & WF3 & F300W & 1300 & 250\nl
4 & WF3 & F300W & 1300 & 250\nl
\nl
5 & WF3 & F336W & 1100 & 0\nl
6 & WF3 & F336W & 1200 & 0\nl
7 & WF3 & F336W & 1200 & 0\nl
8 & WF3 & F336W & 1100 & 250\nl
9 & WF3 & F336W & 1200 & 250\nl
10 & WF3 & F336W & 1200 & 250\nl
\nl
11 & PC & F336W & 700 & 0\nl
12 & PC & F336W & 700 & 0\nl
13 & PC & F336W & 700 & 0\nl
14 & PC & F336W & 800 & 0\nl
15 & PC & F336W & 700 & 0\nl
16 & PC & F336W & 700 & 250\nl
17 & PC & F336W & 800 & 250\nl
18 & PC & F336W & 800 & 250\nl
19 & PC & F336W & 800 & 250\nl
\nl
20 & PC & F300W & 700 & 0\nl
21 & PC & F300W & 800 & 0\nl
22 & PC & F300W & 700 & 0\nl
23 & PC & F300W & 700 & 250\nl
24 & PC & F300W & 800 & 250\nl
25 & PC & F300W & 800 & 250
\enddata
\end{deluxetable}

\clearpage
\begin{deluxetable}{ccccc}
\footnotesize
\tablecaption{Photometric results for each component of lensed quasar
Q2237+0305, in F336W and F300W filters (averaged over all PC 
exposures, cf. Table 1), and positions, obtained from the PC images. 
Photometric quantities are in the 
STMAG system; errors listed are the 1$\sigma$ dispersions of the individual 
results about the mean, added in quadrature with the errors associated
with the zeropoint and with the contamination correction. Astrometric 
quantities are given in arcseconds along the directions of right 
ascension and declination. As explained in the text, the astrometric 
errors do not include possible systematic errors in determining the 
orientation of the coordinate system.} 
\tablewidth{0pt}
\tablehead{\colhead{Component} & \colhead{F336W Magnitude} &
\colhead{F300W Magnitude} & \colhead{Right Ascension (\arcsec)} &
Declination (\arcsec)}
\startdata
A & 16.65 $\pm$ 0.02 & 16.85 $\pm$ 0.03 & 0.000 $\pm$ 0.0015 & 0.000
 $\pm$ 0.0015 \nl
B & 16.93 $\pm$ 0.04 & 17.10 $\pm$ 0.03 & -0.671 $\pm$ 0.0015 & 1.697
 $\pm$ 0.0015  \nl
C & 17.95 $\pm$ 0.03 & 18.24 $\pm$ 0.04 & 0.634 $\pm$ 0.0015 & 1.210 
$\pm$ 0.0015  \nl
D & 18.10 $\pm$ 0.03 & 18.41 $\pm$ 0.03 & -0.866 $\pm$ 0.0015 & 0.528 
$\pm$ 0.0015  \nl
Galaxy Core & N/A & N/A & -0.081 $\pm$ 0.010 & 0.937 $\pm$ 0.010 \\
\enddata
\end{deluxetable}

\clearpage
\begin{figure}
\figurenum{1}
\plotone{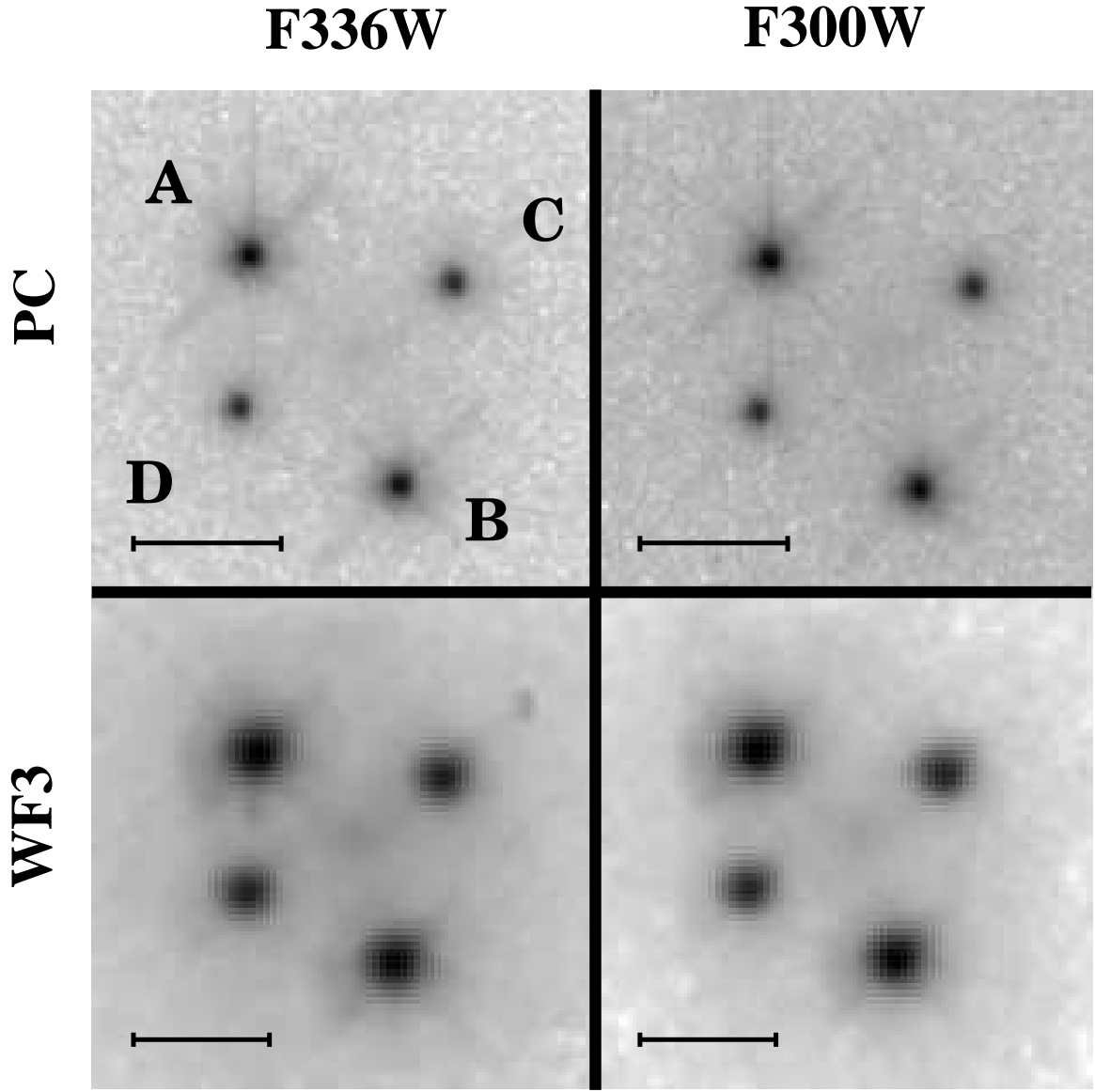}
\epsscale{1.0}
\caption{The full configuration of lensed quasar
Q2237+0305. Components A, B, C, and D are labeled. The images in the
top row were exposed on the PC chip; the scale indicates $1^{''}$.
The images in the bottom row were exposed on the WF3 chip; the scale
indicates $1^{''}$.  The images in the left column used the F336W
filter; those in the right column used the F300W filter. The feature
near component C in the WF3 F336W exposure is a cosmic ray left over
from the image processing.}
\end{figure}

\clearpage
\begin{figure}
\figurenum{2a}
\plotone{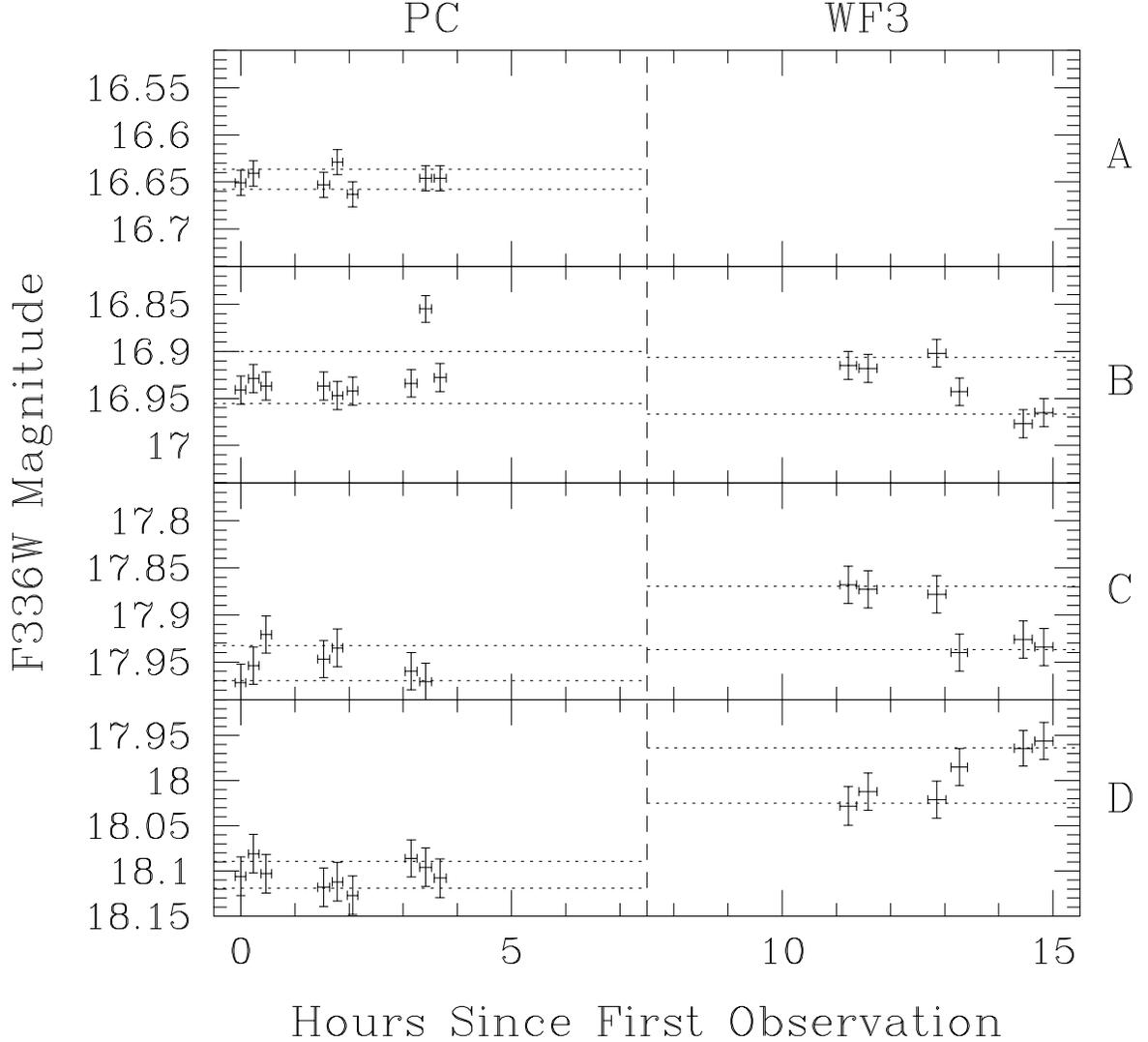}
\epsscale{1.0}
\caption{The ``lightcurve'' of each component of lensed quasar
Q2237+0305 in the F336W filter. The vertical error bars are $1\sigma$
limits determined only from the statistics of the photometry and the
expected pixel-centering error of 0.01 mag; the horizontal error bars
represent the duration of the exposure. Although these images are
corrected for contamination, the error in that contamination
correction is not included in the error bars. Note that the PC chip
was used for the first half of the observations, while the WF3 chip
was used for the second half. Dotted lines delimit the $1\sigma$
dispersion about the mean for the PC and WF3 exposures. Note that the
A component was saturated for the final 8 exposures and therefore we
did not plot those points; other points are missing because of
contamination by cosmic rays.}
\end{figure}

\clearpage
\begin{figure}
\figurenum{2b}
\plotone{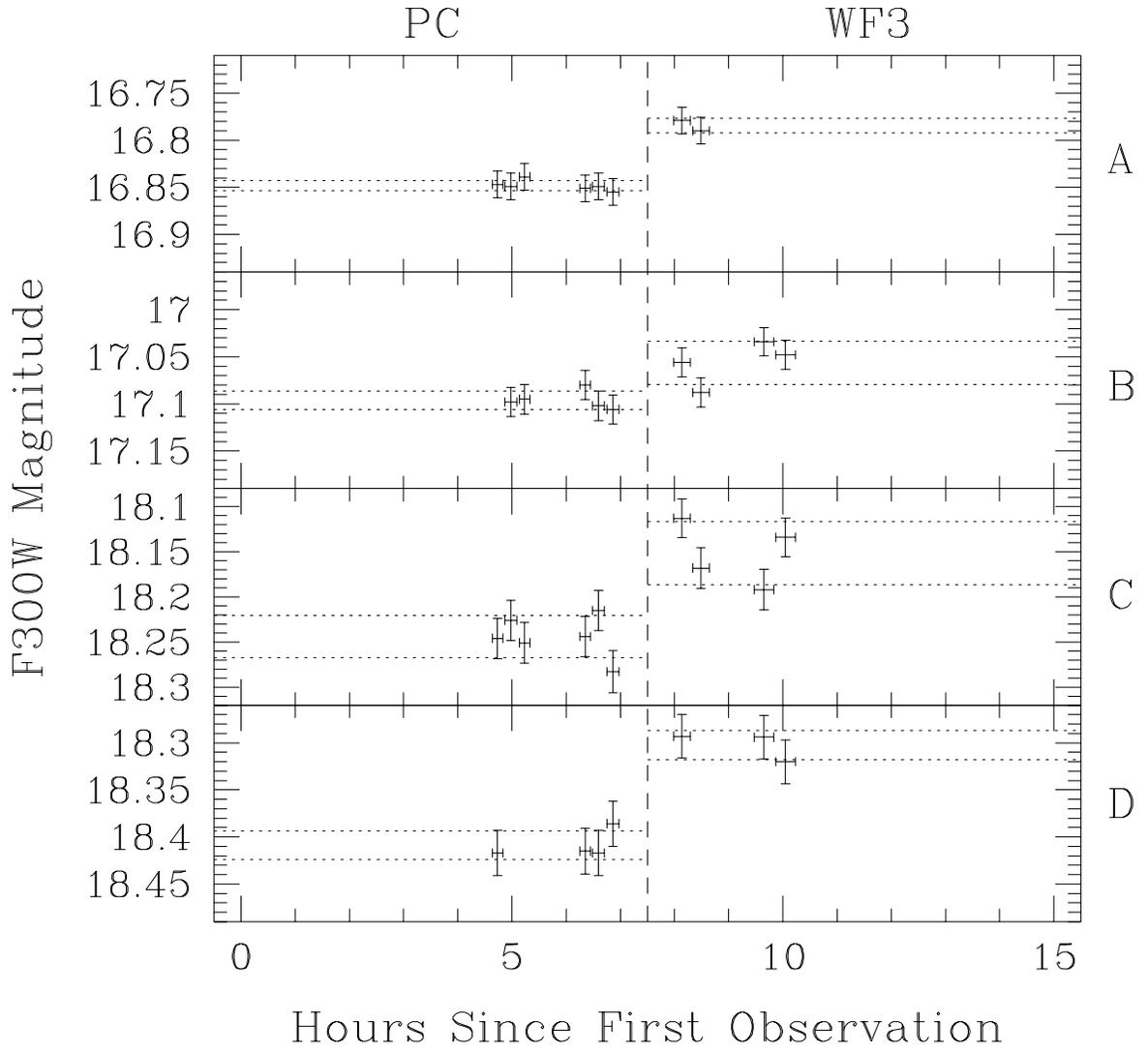}
\epsscale{1.0}
\caption{Same as Figure 2a, for the F300W filter.}
\end{figure}

\clearpage
\begin{figure}
\figurenum{3}
\plotone{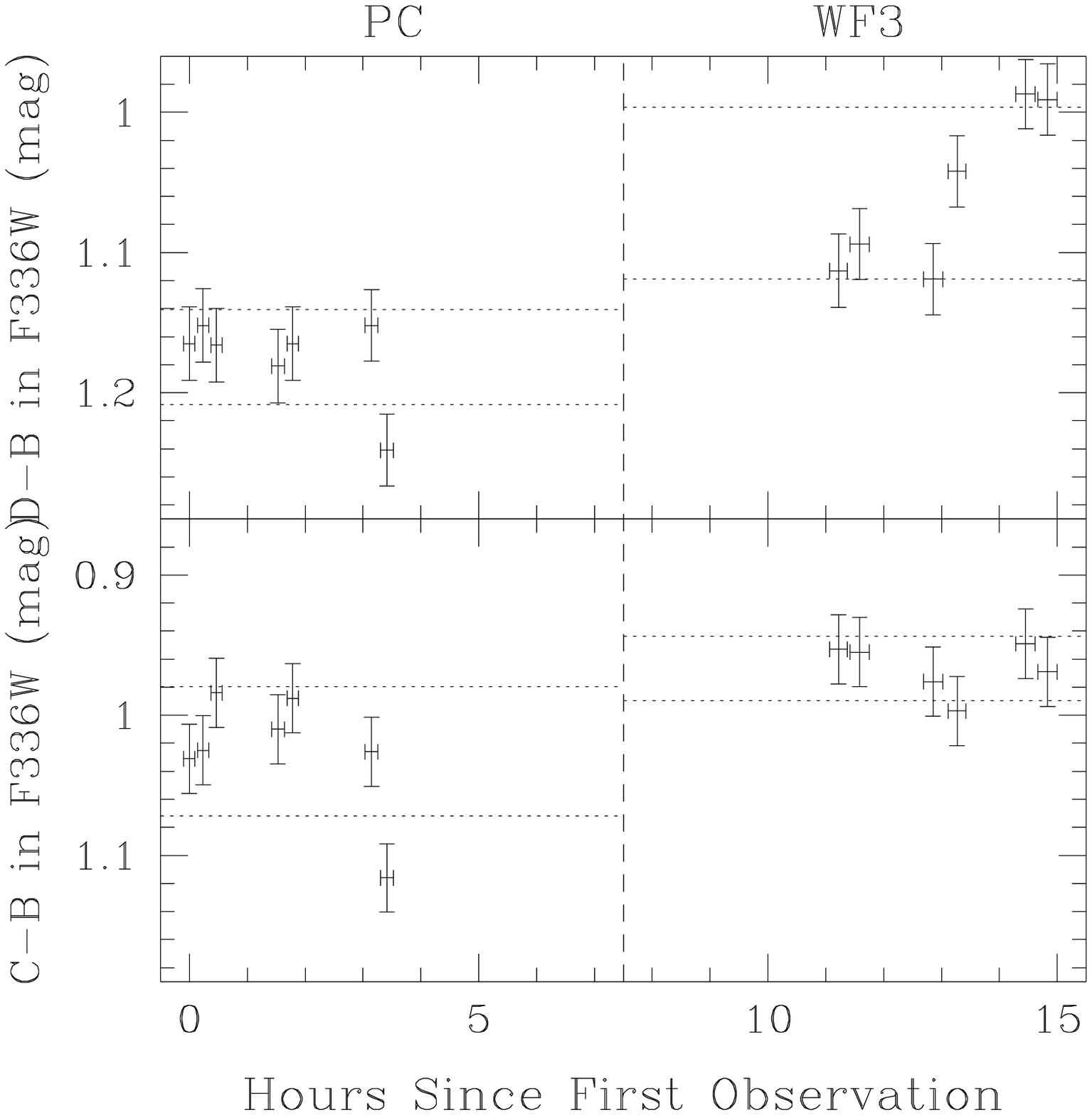}
\epsscale{1.0}
\caption{D$-$B and C$-$B magnitude differences for lensed quasar
Q2237+0305 in the F336W filter. The vertical error bars are $1\sigma$
limits determined only from the statistics of the photometry and the
expected pixel-centering error of 0.01 mag; the horizontal error bars
represent the duration of the exposure. Dotted lines delimit the
$1\sigma$ dispersion in the D$-$B and C$-$B values about their mean for
the PC and WF3 exposures. Note the rise in D$-$B in the final three
images.}
\end{figure}

\clearpage
\begin{figure}
\figurenum{4}
\plotone{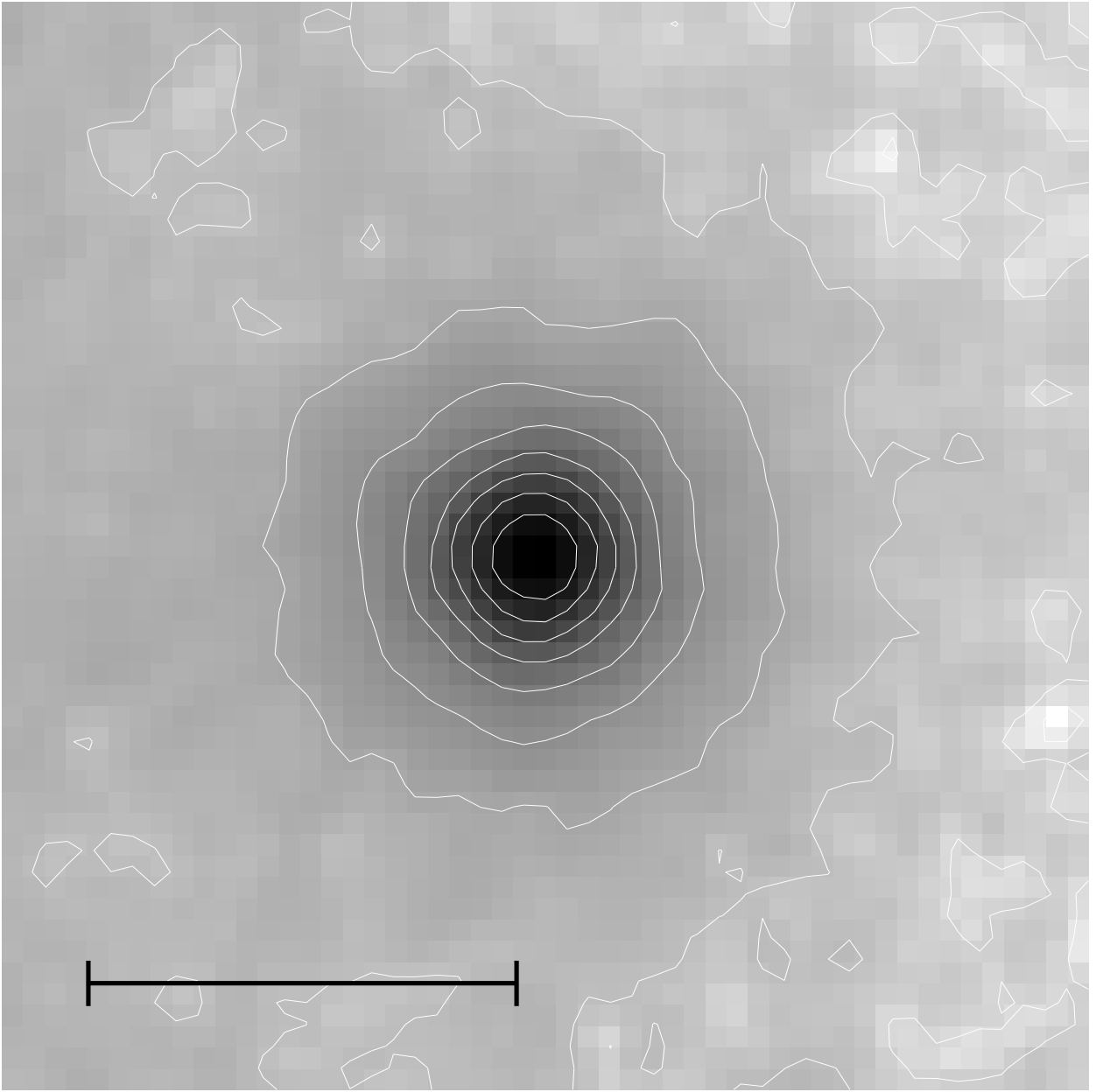}
\epsscale{1.0}
\caption{A sum of all the components of lensed quasar Q2237+0305 from
the images taken in the PC chip with the F336W filter. We aligned the
images of each component such that the direction to the galaxy is to
the left of the page and the clockwise direction around the galaxy is
down the page. The contours reveal about a 5\% extension in the
y-direction, which we believe to be consistent with no extension. The
scale indicates $0.5^{''}$.}
\end{figure}

\clearpage
\begin{figure}
\figurenum{5a}
\plotone{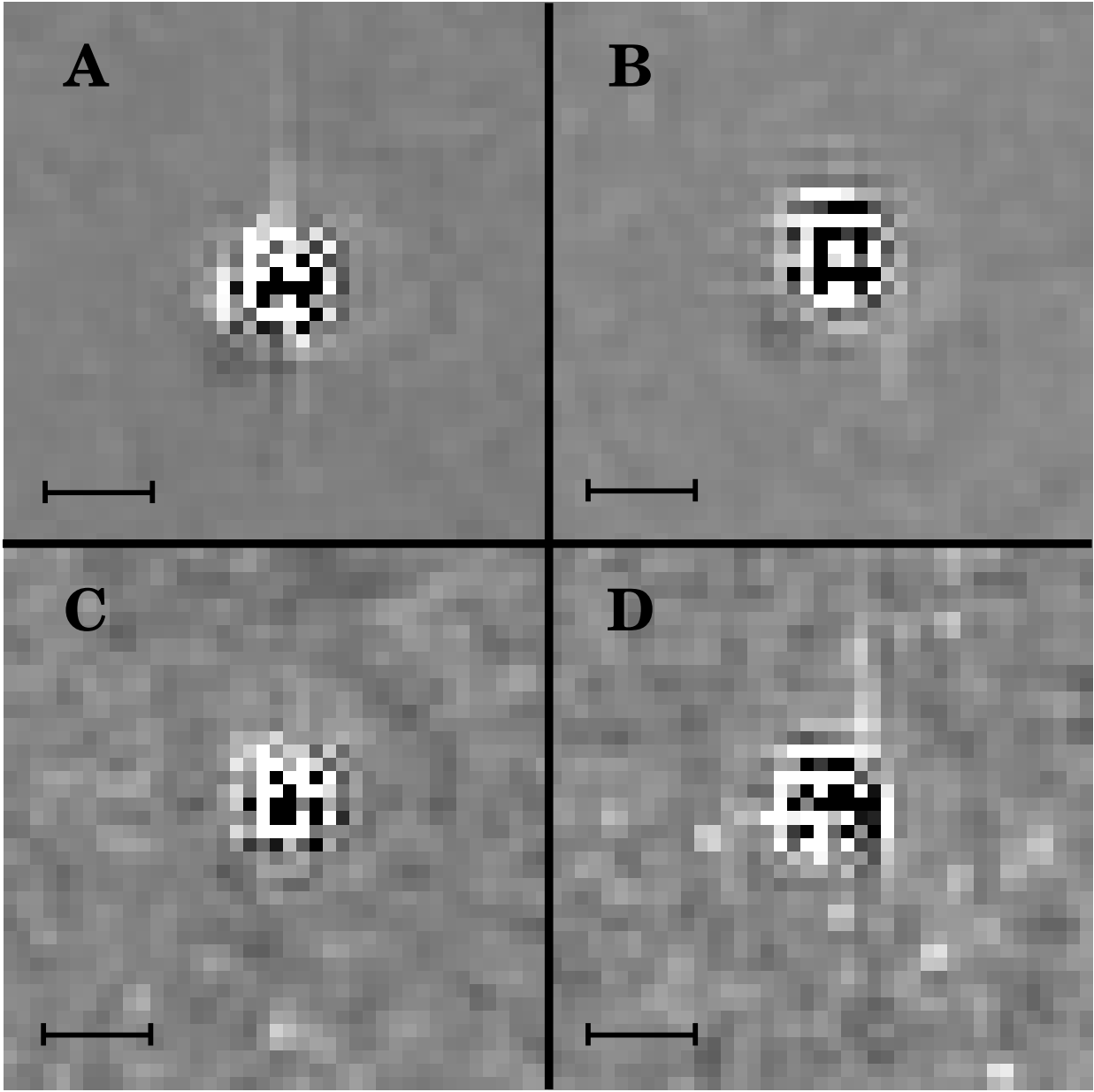}
\epsscale{1.0}
\caption{The scaled differences between the F336W and F300W filter in
the PC chip for each component of lensed quasar Q2237+0305. Darker
colors indicate more intense F336W. Note the features on the lower left of
components A and B. The scale indicates $0.2^{''}$.}
\end{figure}

\clearpage
\begin{figure}
\figurenum{5b}
\plotone{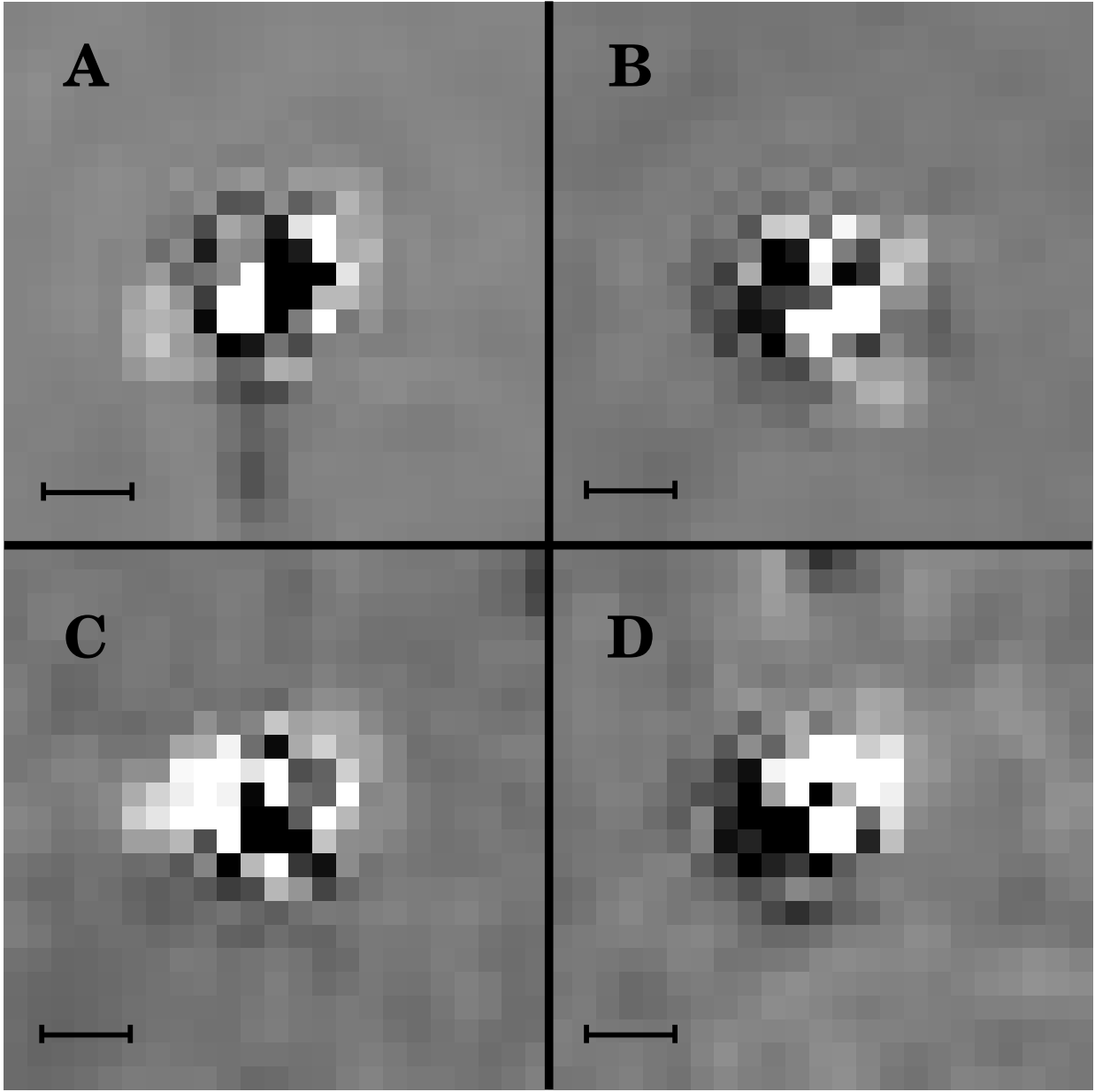}
\epsscale{1.0}
\caption{Same as Figure 3a, but for the WF3 chip. The scale indicates
$0.2^{''}$.}
\end{figure}

\clearpage
\begin{figure}
\figurenum{6}
\plotone{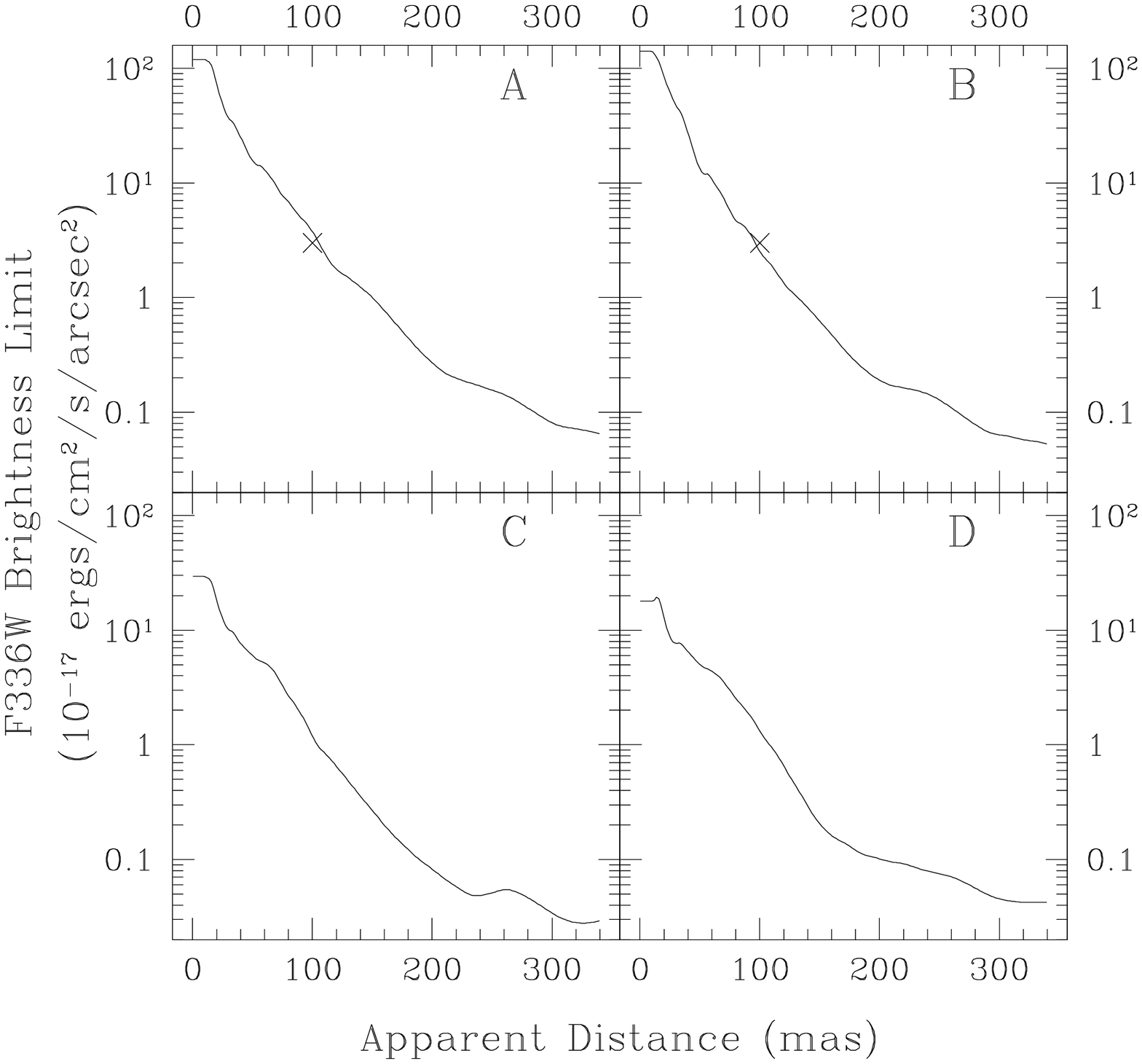}
\epsscale{1.0}
\caption{Limits on the brightness of a possible Lyman-$\alpha$ region,
according to the PC chip in the F336W filter. Apparent distance is the
angular distance measured on the sky. It equals the angular distance
in the source plane multiplied by the magnification. The plotted
curves correspond to $1\sigma$ limits. Crosses in the A and B plots
indicate where the feature of Figure 3a lies.}
\end{figure}

\clearpage
\begin{figure}
\figurenum{7}
\plotone{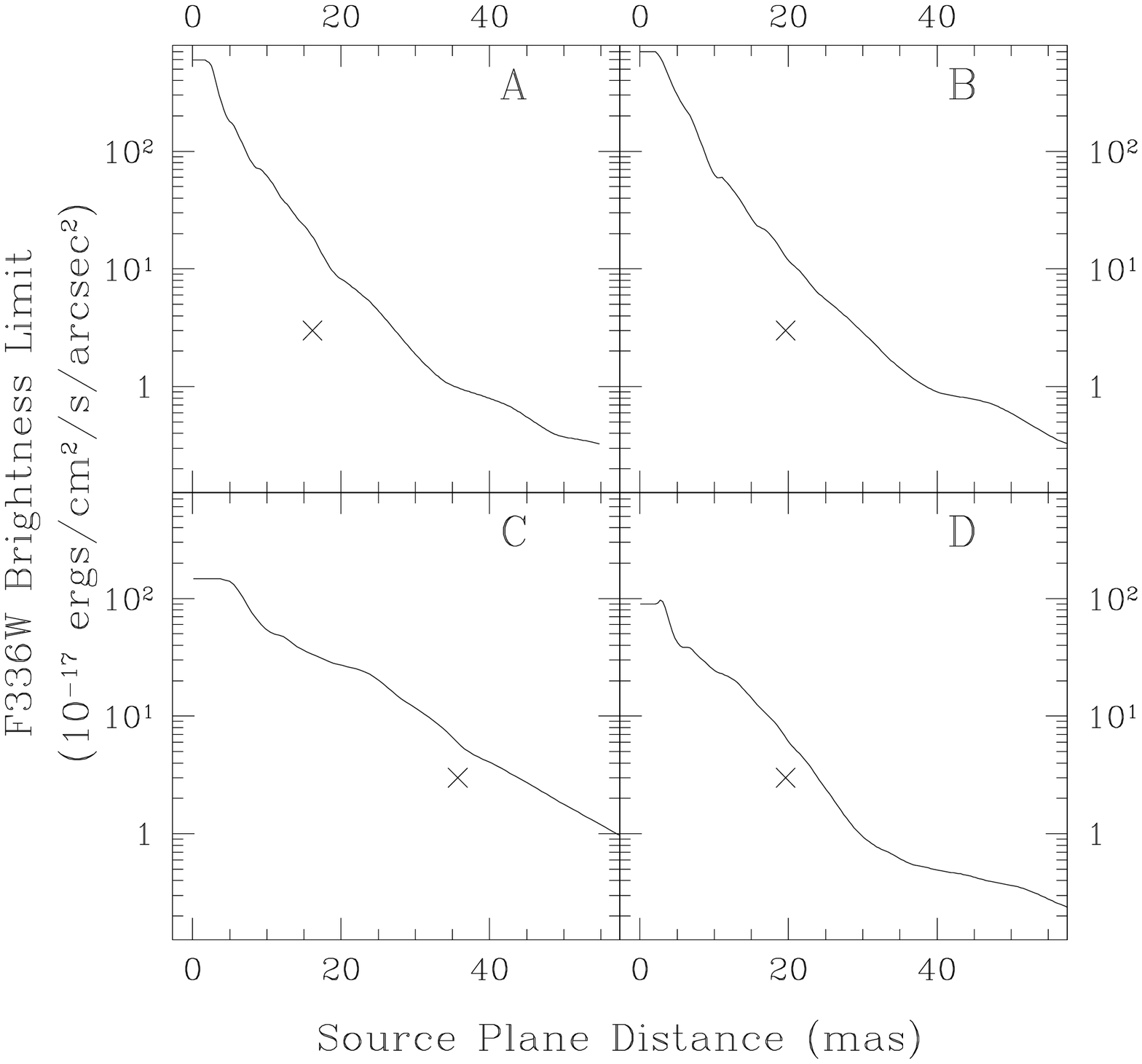}
\epsscale{1.0}
\caption{Limits on the brightness in the source plane of a possible
Lyman-$\alpha$ region, according to the PC chip in the F336W
filter. Magnifications are taken from the model of Rix (1992), and the
curves correspond to a 5$\sigma$ result. Crosses indicate where the
feature of Figure 3a would lie in the source plane; we determined its
brightness and its position in the source plane by examining its
incarnation next to the B component.}
\end{figure}

\end{document}